**Title:** Sublinear Matching With Finite Automata Using Reverse Suffix Scanning

**Author**: Steven Kearns[1]

1: Software Truth
2818 Juniper St.
San Diego, CA 92104
United States
skearns23@yahoo.com / 858-603-8828

**Corresponding Author**: Steven Kearns

**Abstract**:

We give algorithms to accelerate the computation of deterministic finite automata (DFA) by calculating the state of a DFA n positions ahead utilizing a reverse scan of the next n characters. Often this requires scanning fewer than n characters resulting in a fraction of the input being skipped and a commensurate increase in processing speed. The skipped fraction is > 80% in several of our examples. We introduce offsetting finite automata (OFA) to encode the accelerated computation. OFA generalize DFA by adding an integer offset to the current input index at each state transition. We give algorithms for constructing an OFA that accepts the same language as a DFA while possibly skipping input, and for matching with an OFA. Compared to previous algorithms that attempt to skip some of the input, the new matching algorithm can skip more often and can skip farther. In the worst case the new matching algorithm scans the same number of characters as a simple forward scan, whereas previous approaches often scan more, so the new algorithm can be used as a reliable replacement for the simple forward scan. Additionally, the new algorithm adapts to available memory and time constraints.

**Biography of Steven Kearns:**

Steven Kearns received a Ph.D. in Computer Science from Columbia University in 1991 where his thesis showed how to implement extended regular expressions that include context sensitive operators, and also gave efficient algorithms for extracting a full parse from a regular expression match. He currently works as a software consultant primarily for the financial industry.

## 1 Introduction

We can efficiently test whether a regular expression $e$ of length $m$ matches a string $s$ of length $n$ by first compiling the regular expression into an equivalent deterministic finite automaton (DFA) $M$, and then running $M$ on $s$; if the resulting state of $M$ is a final state, then the regular expression matches $s$. The matching process ("running $M$ on $s$") requires O($n$) time [1].

There are other approaches that require less time to prepare a regular expression for matching but more time to do the actual matching. For instance, a regular expression can quickly be converted to a nondeterministic finite automaton (NFA), which is then used for matching by simulating all possibilities in parallel; this approach typically requires O($m$) time for computing the equivalent NFA, O($m$) space for storing the NFA, and O($mn$) time for matching. Another alternative is "lazy DFA compilation" which offers a balance between fast compilation and fast matching in most cases.

The algorithm in this paper is most appropriate when the matching time dominates the preprocessing time, which means that either the regular expression of interest has an equivalent DFA of small or reasonable size, or the total input to be matched is very long.

*Sublinear matching* is locating a pattern in an input string without examining every character. Although typically a sublinear algorithm is worst-case equivalent to an algorithm that examines every character, in practice these algorithms run faster because the execution time is proportional to the number of characters examined.

Sublinear matching was pioneered in the Boyer-Moore string matching algorithm [2], where it was used to find matches to a single string *s1* within a longer string *s* without examining all of the characters of *s*. This algorithm has since been improved to handle sets of strings and to skip more often and farther; [3] has an overview.

Some applications require that we find all substrings of input string *s* which match a regular expression *e*. We will call this problem *FindSubstrings(s, e)*. Since there are $(n+1)^2$ different substrings when *s* has length *n*, the worst case complexity of any algorithm for this problem is $O(n^2)$ at best. [4] gave a Boyer-Moore inspired algorithm for *FindSubstrings(s, e)*.

Because of the $O(n^2)$ worst case behavior of algorithms for *FindSubstrings(s, e)*, many applications instead solve the problem "find all positions in *s* marking the end of a match to regular expression *e*". This problem, which we call *FindEndPositions(s, e)*, can be solved in $O(n)$ (ignoring the time to create the DFA). The technique is to create regular expression *ee* where $ee = (\sum^* e)$ which matches any string whose suffix matches *e*; then we create a DFA *FF* equivalent to *ee* and run *FF* on *s* using a left to right scan of *s*. The DFA will enter a final state at every position of *s* marking the end of a match to *ee*, and thus the end of a match to *e*. If we need to find the full substring matching a regular expression, or we need to find a full parse of a regular expression match, we can use the rightmost position of a match and the parsing techniques of [5] or [6]. Or, we can use a backwards scan using a DFA or NFA equivalent to *reverse*(*e*) from the rightmost position [5].

Three algorithms for solving *FindEndPositions(s, e)* are presented in [7] that can skip characters. Two of the algorithms provide worst case $O(n)$ matching. These algorithms use both forward and backward scans, and are able to avoid examining some characters in the input when a "bad substring" is identified during the backward scan. A "bad substring" is one that cannot be part of any match to e. If found the algorithms jump to a subsequent position marking a possible start of a match. Unfortunately, the algorithms may examine some characters twice, once in a backward scan and again in a forward scan. Also, the algorithms are more complex than the forward-only algorithm. As a result, the algorithms often perform slower than the forward-only algorithm.

The new algorithm reported here solves *FindEndPositions(s, e)*, and is applicable to the pure DFA approach. We do more preprocessing but get faster matching by skipping more often and farther. Our algorithm never processes a substring more than once and may avoid processing some substrings altogether. In addition, it operates very similarly to the traditional forward scan algorithm, so these enhanced capabilities do not incur significant overhead. Therefore, for applications that can spare the expense of greater preprocessing the new algorithm is a reliable replacement the traditional algorithm. It will match no worse, and usually faster, than either the traditional forward scan algorithm or the algorithm in [7].

**2 Basic Definitions**

Regular expressions are patterns composed using characters from an alphabet $\sum$, the sequence operator (p1 p2), alternation (p1 | p2), and 0-or-more repetitions p1* or 1-or-more repetitions p1+. See [1] for the standard introduction.

We assume a string type with the usual operators. *len*(*s*) is the length of string *s*, and the characters of *s* are indexed from 0 to *len*(*s*)−1 inclusive and accessed as *s*[*i*]. *rev*(*s*) is the reverse of *s*. *last*(*s*) = *s*[*len*(*s*)−1] is the last character in *s*. *s*+*t* for string *s* and string or character *t* is the concatenation of *s* and *t*.

String positions fall between string characters: position *i* is before *s*[*i*] and after *s*[*i*−1]. Position *len*(*s*) is at the end of *s*. Substrings of *s* are written *s*[*i*..*j*] and represent the sequence of characters of length (*j*−*i*) between positions *i* and *j*. *s*[*i*..*i*] is the empty string. *prefix*(*s*, *i*) is *s*[0..*i*] and is the length *i* prefix of *s*.

A deterministic finite automaton (DFA) has a set of states $Q$, a set of final states $F \subseteq Q$, a distinguished starting state $q0$, an alphabet $\Sigma$, and a transition function $\delta(q,c)$ of type $(Q \times \Sigma) \rightarrow Q$. When we are discussing multiple automata such as M1 and M2 we may write $Q_{M1}$, $F_{M1}$, etc. to distinguish the components of the different automata.

The transition function $\delta(q,c)$ is extended to a transition function $\delta(q,s)$ defined on strings in the standard way: $\delta(q,\varepsilon) = q$ and $\delta(q,c+t) = \delta(\delta(q,c), t)$. A DFA *matches s* when $\delta(q0, s) \in F$.

## 3 Reverse Suffix Scanning: Concepts and an Example

Reverse suffix scanning is based on the observation that $\delta(q, s)$ is often determined by a proper suffix of *s*. Figure 1 is a graphical example of a DFA that illustrates this possibility.

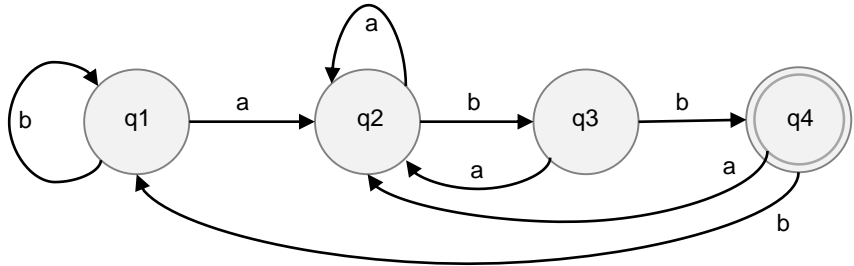

**Figure 1:** DFA equivalent to (a|b)*(a b b)+

To compute $\delta(q1, s)$ for *s* of length 3, we note that if *s* ends in "a" then the result will be q2 irrespective of the first two characters of *s*. However, if *s* ends in "b" then we cannot determine the resulting state unless we examine other characters in *s*, because for instance "abb" results in state q4 while "aab" results in q3.

### 3.1 Building a Reverse Scanning Trie

This suggests that to compute $\delta(q1, s)$ for *len*(*s*) = 3 while scanning *rev*(*s*) we tabulate a map from all possible *rev*(*s*) to $\delta(q1, s)$. The choice of len(s) = 3 is arbitrary, within some constraints discussed later, so we generalize this value as *look*(q1) and note that every state can have a different lookahead value.

| | | | |
|---|---|---|---|
| aaa → q2 | aba → q2 | bba → q4 | baa → q3 |
| aab → q2 | abb → q2 | bbb → q1 | bab → q3 |

To efficiently store and lookup the possibilities we use a trie where each path from the root to a leaf is *rev*(*s*), which of course means that the path from that leaf to the root is *s*, and whose leaves are labeled with the state $\delta(q1, s)$. Figure 2 has a trie of depth 3 that encodes the same mapping.

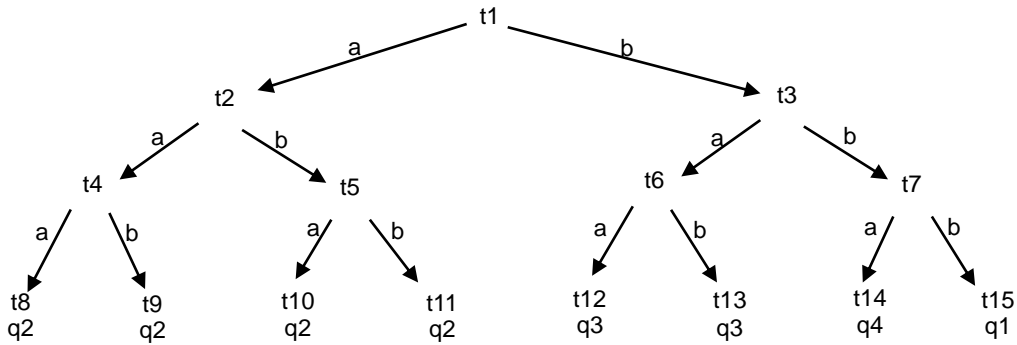

**Figure 2:** Trie for q1

To use this trie to compute $\delta$(q1, "aab"), we start at t1 and follow the branches corresponding to *rev*("aab"), or equivalently we read the characters of "aab" in reverse order "baa". We transition from t1 → t3 on "b", then t3 → t6 on "a", then t6 → t12 on "a". Since we reach a leaf, we are done and the resulting state is q3.

To formalize this process we give some basic definitions for tries. In the following, assume that *T1*, *T2* are tries; *q* is a state of a DFA; *lookahead* is a positive integer; *c* and *cx* are characters; and *s* is a string.

Tries are recursive unbalanced tree data structures. A trie *T1* can be either a leaf or a trie node. If *T1* is a leaf then *isLeaf*(*T1*) is true, it has no children, and *state*(*T1*) is a state of an associated finite automaton. If *T1* is a trie node then *isNode*(*T1*) is true and *child*(*T1*, *c*) is the child trie indexed by character *c* and *offset*(*T1*, *c*) is an integer offset indexed by *c*. All trie node *T1* are complete, which means that *child*(*T1*, *c*) has a value for every *c* in the alphabet. *level*(*T2*) of a trie is the distance of *T2* from the root of the trie, which is 0 at the root. In our algorithms we create a leaf via the function *trieLeaf*(*q*). We create a trie node with no children using the function *trieNode*(). We set *child*(*T1*, *c*) by writing "*child*(t, *c*) = ...", and we set *offset*(*T1*, *c*) similarly.

Definition 1 describes how *selectState*(*T1*, *s*) uses *T1* to map *s* to a result state by scanning *s* in reverse and traversing from the root of *T1* to a leaf.

**Definition 1**: *selectState*(*T1*, *s*) =
      if *isLeaf*(*T1*) then *state*(*T1*) else *selectState*(*child*(*T1*, *last*(*s*)), *prefix*(*s*, *len*(*s*)−1)) fi
where *len*(*s*) ≥ *depth*(*T1*)

Definition 2 describes what state *selectState*(*T1*, *s*) returns.

**Definition 2**: *isTrieForState*(*T1*, *q*, *lookahead*) ⇔ (*selectState*(*T1*, *s*) = $\delta$(*q*, *s*) for all *s* having *len*(*s*) = *lookahead*).

To solve *FindEndPositions*(*s*, e) we need to locate all positions *p* in *s* where $\delta$(*q*, *s*[0..*p*]) is a final state. Our key tool in this effort is Lemma 3 which describes how to find the state at a position after *p* if we know the state at *p*. Lemma 3 follows directly from from Definition 1 and Definition 2 so we omit its proof.

**Lemma 3:** If we are in state *q* at position *p* of *s* and *isTrieForState*(*T1*, *q*, *look*(*q*)) then *selectState*(*T1*, *s*) is the state at position *p* + *look*(*q*). Also, *selectState*(*T1*, *s*) works by scanning a suffix of *s*[*p*..*p* + *look*(*q*)] in reverse order. Therefore the first character processed is at index *p* + *look*(*q*) − 1.

Why did we pick *look*(q1) = 3? Our goal is to identify *all* positions of an input string where we are in a final state. Our trie of lookahead 3 computes the state after processing 3 characters, but it does not compute the intermediate states, for instance the state after 2 characters. To ensure that none of the skipped intermediate states are final states, we ensure that no final state is closer than 3 transitions away. Applying this logic, we see that the max *look*(q1) and *look*(q4) is 3, max *look*(q2) is 2, and max *look*(q3) is 1. In general, max *look*(*q*) is *finalDist*(*q*) which we now define as the distance from *q* to a final state.

**Definition 4:** *finalDist*(*q*) for state *q* is the minimum length string *s* for which *len*(*s*) > 0 and $\delta$(*q*,*s*) ∈ F. (The exclusion of the empty string is perhaps non-standard but is most appropriate for our use.)

*3.2 Compressing a Trie*

So far, what we have done is implement a nice way to compute $\delta$(q1, *s*) by reading the characters of *s* in reverse order. Next we "compress" the trie by taking advantage of the fact that if all children of a node are leaf nodes labeled with the same state *q*, then we can convert the node to a leaf node labeled with *q*. Here are the formal definitions:

**Definition 5:** *compressible*(*T1*) ⇔ (*isNode*(*T1*) and all children are leaves that have the same state value).

**Definition 6:** *compressNode*(*T1, T2*) returns a *T1'* which is the same as *T1* except some descendant node *T2* for which *compressible*(*T2*) is replaced with *T2'* = *child*(*T2, c*) for arbitrary *c*. (Since all children are leaves with the same state, *child*(*T2, c*) returns an equivalent leaf for any *c*.)

**Definition 7:** *compressed*(*T1*) ⇔ (no descendant of *T1* is compressible).

**Definition 8:** *subRootCompressed*(*T1*) ⇔ (*isNode*(*T1*) and no proper descendants of *T1* is compressible). This differs from *compressed*(*T1*) in that *subRootCompressed*(*T1*) allows the possibility that the root of *T1* remains compressible but not compressed.

Clearly we can establish compressed(T1) by repeatedly locating a compressible node T2 in T1 and updating T1 to compressNode(T1, T2). An example of a compressible node is t4: each child of t4 is a leaf whose state is q2. Equivalently, because aaa → q2 and aab → q2 we can say aaX → q2 where "X" means "dont care". Applying this rule to all nodes, bottom up, we ensure the trie is compressed. Our example trie of Figure 2 becomes the compressed trie in Figure 3. (The numbers accompanying each transition will be explained shortly.)

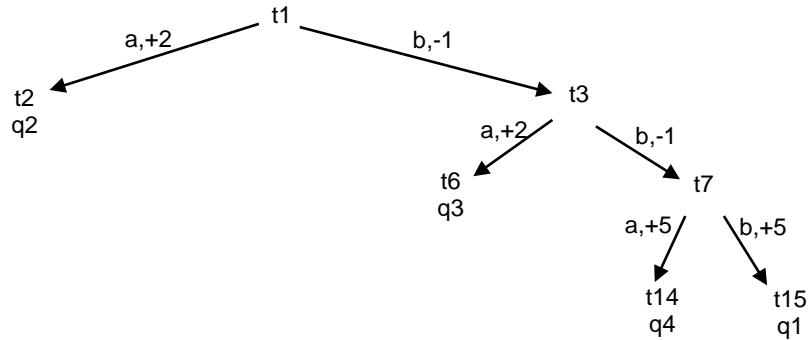

**Figure 3**: Compressed trie for q1 and q4

After compressing the trie, some paths from the root to a leaf will have length < 3, which is great because it means that in some cases we can compute $\delta(q1, s)$ while reading a proper suffix of *s* in reverse order. Returning to our earlier example, we now see that we can compute $\delta(q1, \text{"aab"})$ by following the branches corresponding to *rev*("aab") in which case we transition from t1 → t3 → t6 on "ba", and we have the answer without having to read the final "a" at all.

Clearly, to achieve a maximal compression and as a result maximize opportunities to skip characters during matching, one must employ DFA minimization before building tries. The reason is that DFA minimization may make states equal which were formerly different, and as a result a node may become compressible because leaves with different states now have equal states.

Note that a trie *T1* which is maximally compressed might result in *T1* being a leaf. This causes complications when we convert our tries to an OFA (described in Section 3.5). To simplify the exposition we stipulate *subRootCompressed*(*T1*) in the tries we build

We are obligated to show that compressing a trie maintains its crucial property, *isTrieForState*(*T1, q, lookahead*), which Theorem 9 states and proves.

**Theorem 9**: *isTrieForState*(*T1, q, lookahead*) implies *isTrieForState*(*compressNode*(*T1, T2*), *q, lookahead*) when *T1* has a compressible descendant T2.

**Proof**: Let *T1'* = *compressNode*(*T1, T2*). We will show that *selectState*(*T1', s*) = *selectState*(*T1, s*) for all *s* where *len*(*s*) = *lookahead*, because then the theorem is true from Definition 2. Since *T1* and *T1'* differ only at and below *T2*, the only interesting case is when *s* causes a traversal to *T2*. In this case, *selectState*(*T1, s*) will continue traversing to a child leaf of *T2* and return *q2*, where *q2* is the state value common to all children of *T2*. Also, *selectState*(*T1'*, *s*) will traverse to *T2'* and then return *q2* since *T2'* is a leaf equal to *child*(*T2*, *c*) which has state *q2*. Therefore we see *selectState*(*T1*, *s*) = *selectState*(*T1'*, *s*) = *q2* when *s* traverses to *T2*. □

*3.3 Matching With Reverse Scanning Tries*

We will be computing a trie for every state of the DFA. For instance, there will be a trie for q4 which happens to look exactly like the trie for q1 except nodes t1, t3, t7, .. are renamed to r1, r3, r7... **Figure 4** and **Figure 5** are the compressed tries for q2 (lookahead 2) and q3 (lookahead 1).

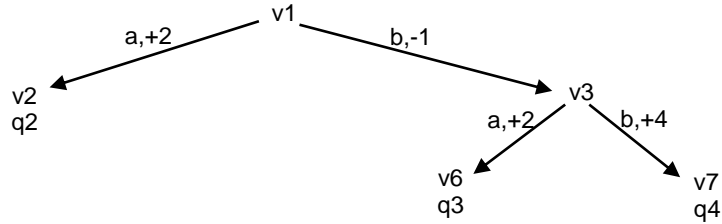

**Figure 4:** Trie for q2

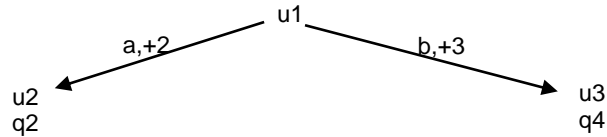

**Figure 5:** Trie for q3

We show later that the worst case time and space to build a trie for a state q is exponential: $O(|\Sigma|^{look(q)})$. It may seem unreasonable to build a trie that can grow exponentially large for each state of a DFA when the number of states of a DFA is worst case exponential in the size of the equivalent regular expression. However, we can always choose to build a trie of lookahead 1 for any state *q*, and such a trie is equivalent to the transition function $\delta(q, c)$. This means it requires comparable space and time to build a trie for q with lookahead 1 as is required in the original DFA. Since most states of a DFA are rarely used during matching we have the option of building larger tries only for states that offer the best chance of skipping characters, assuming we can identify such states. Also, we will later give an algorithm for growing tries by increasing their lookahead by l at each iteration. This makes it easy to limit the size of any trie to a reasonable value.

To solve the *FindEndPositions*(*s*, *e*) problem we need to find all positions in *s* where a DFA related to *e* is in a final state. We describe how we use tries to accomplish this task, assuming we have a trie for each state of the DFA. Since the DFA will start in q0 at position 0 of *s*, we can use the trie for q0 and apply Lemma 3 to calculate the state, call it q2, at position $p2 = 0 + look(q0)$. Then we repeat the process, using the trie for q2 to calculate the state at position $p3 = p2 + look(q2)$. We iterate this process until the input is exhausted.

Figure 6 shows a trace of a match using our example tries to match "abbabaabb..." We start in q1 at position 0; then we use the trie for q1 on *rev*(*s*[0..3]) to compute q4 at position 3; then we use the trie for q4 on *rev*(*s*[3..6]) to compute q2 at position 6; etc. The arrows show the sequence of characters inspected during the matching process, and the sequence of trie nodes traversed, starting at root t1 of the trie for initial state q1 and index 2. Note that we were able to skip 2 characters out of the first 9 in this example. Also note we discovered two positions, 3 and 9, where the state, q4, was a final state, which means that *s*[0..3] and *s*[0..9] are strings accepted by the example DFA in Figure 1. Finally, note that we do not compute the state at every position.

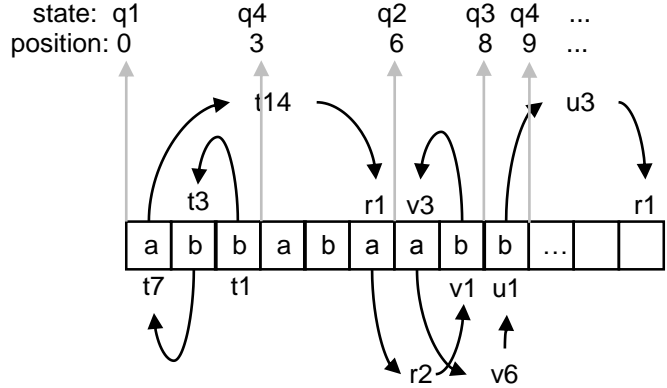

**Figure 6:** Trace of the matching algorithm

*3.4 Assigning Offsets*

Now we can describe the numbers that label each transition in our tries. These numbers give the offset from the current index in the string to the next index to process. For any transition from a trie node to a non-leaf, this offset will be −1 because the trie for state $q$ computes $\delta(q, s[i..j])$ by processing $s[j−1]$, $s[j−2]$, ... in turn. For any transition from a trie node T2 to a leaf node T3 the offset will be $level(T2) + look(state(T3))$. The factor $level(T2)$ returns the index to its starting point for the trie T2 and the factor $look(state(T3))$ additionally offsets the index to the first character to read for the next trie. For example, in Figure 4 the transition from v1 to v3 is from a node to a node, so the offset is −1. The transition from v3 to v7 is +4 because v7 is a leaf and $level(v3) = 1$ and $look(state(v7)) = 3$.

**Definition 10**: AssignOffsets(*T1*) sets *offset*(...) for each transition in *T1*. For all trie node *T2* in *T1* and *c* in $\sum$, let *T3* = *child*(*T2*, *c*) in *offset*(*T2*, *c*) = if *isNode*(*T3*) then −1 else *level*(*T2*) + *look*(*state*(*T3*)) fi.

*3.5 Converting to an Offsetting Finite Automaton*

As presented, after we traverse to a leaf node we use the state at the leaf to select the next trie. We get an interesting perspective if we replace all transitions to a leaf node *T2* with a transition directly to the trie for *state*(*T2*). Definition 11 details the transformation, and Figure 7 shows the result. Note that after calling *LinkTries*(*T1*) on every trie, all leaves have been removed.

**Definition 11**: *LinkTries*(*T1*) replaces all transitions in *T1* to a leaf node with a transition to the trie for the state at the leaf node. For all trie node *T2* in *T1* and *c* in $\sum$ where *isLeaf*(*child*(*T2*, *c*)), set *child*(*T2*, *c*) = *trieForState*[*state*(*child*(*T2*, *c*))].

To relate the combined trie to our previous tries, note that nodes t1, t3, t7 are from the trie for q1; nodes r1, r3, r7 are from the trie for q4; nodes v1, v3 are from the trie for q2; and node u1 is from the trie for q3. Interestingly, our combined tries now looks like a finite automaton where each transition is labeled with an offset in addition to the usual character. At each transition the automaton also maps an index to a next index by adding an offset. The non-leaf nodes of the trie for each state become states of the new automaton. The root nodes of the tries for the initial and final states (t1 and r1) serve the same roles in the new automaton. We call this automaton an Offsetting Finite Automaton (OFA). We are interested in OFA because it is a minimal extension of DFA needed to describe a matching process that can scan input characters in a non linear order.

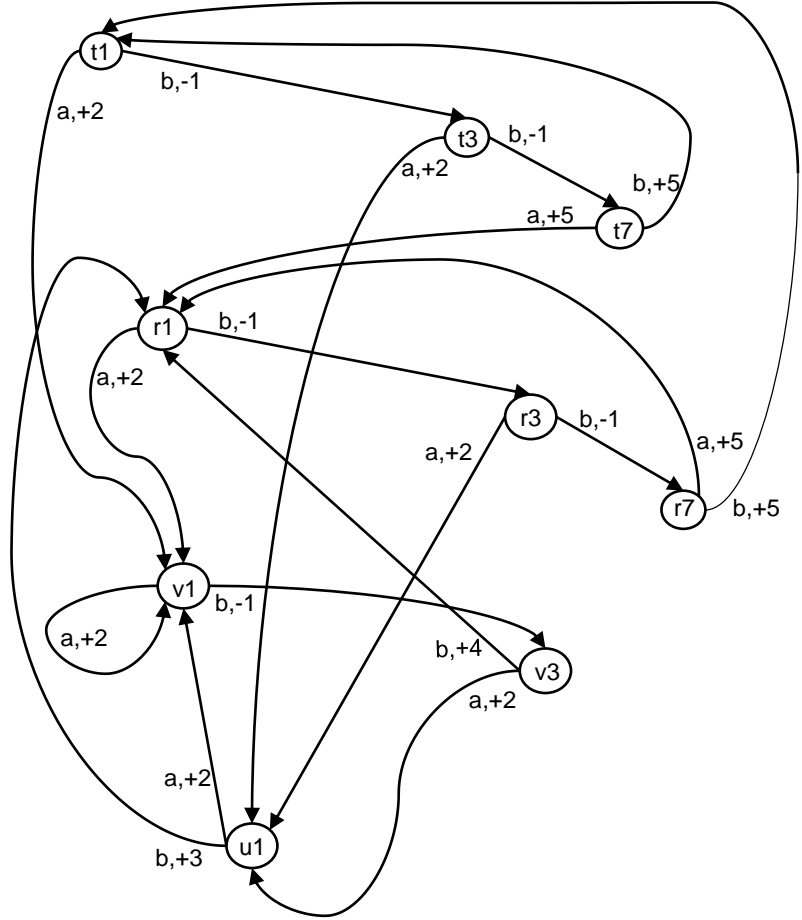

**Figure 7**: Combined tries, directly linked

**Definition 12**: An Offsetting Finite Automaton (OFA) has a set of states $Q$, a set of final states $F \subseteq Q$, a distinguished initial state $q0$, a state transition function $\delta(q,c)$ of type $(Q \times \Sigma) \rightarrow Q$, an indexOffset function $\Theta(q,c)$ of type $(Q \times \Sigma) \rightarrow \mathbb{Z}$, and a positionOffset function $\Phi(q)$ of type $(F \cup q0) \rightarrow \mathbb{Z}$. At each transition an OFA maps a tuple $<q, i> \in <Q, \mathbb{Z}>$ to a new tuple $< \delta(q,c), i+ \Theta(q,c) >$. The $i$ tuple component represents the index of the next character to process. We define the action $\Delta$ of OFA on a string $s$ by iterating while $i$ is a valid index:

$\Delta(<q, i>, s) =$ if $0 \leq i < len(s)$ then $\Delta(< \delta(q, s[i]), i + \Theta(q, s[i]) >, s)$ else $<q, i>$ fi

We say that an OFA matches $s$ when

$<q', i'> = \Delta(<q0, \Phi(q0)>, s)$ and $q' \in F$ and $i' - \Phi(q') = len(s)$

We explain the role of positionOffset $\Phi(q)$ and how it differs from indexOffset $\Theta(q, c)$. During matching we typically compute "the state at position $p$ of $s$" for increasing $p$, looking for positions where the state is a final state. For DFA, if we are in state $q$ at position $p$ of string $s$ then the next character to process is at index $p$. Therefore for DFA the "current position", call it $cp$, is always equal to the "next index to process", call it $ni$. However, for OFA the situation is more complicated. The rule for OFA is $cp + \Phi(q) = ni$. This is relevant when we start matching, because if we start in initial state $q0$ at position 0, then the next character to process is at index $0 + \Phi(q0)$. It is also relevant when we find ourselves in a final state $q$ with next index to process $i$ because then we can conclude that we are in state $q$ at position $i - \Phi(q)$. The positionOffset function is otherwise irrelevant because the index $i$ in $\Delta(<q, i>, c)$ always represents the next index to process which is updated with the indexOffset function. This also explains why the OFA definition requires $\Phi(q)$ to be defined only on the initial and final states.

The value of $\Phi(q)$ is easy to calculate from Lemma 3: if we are in state $q$ at position $p$ then the next character to process is at index $p + look(q) - 1$, therefore we have $\Phi(q) = look(q) - 1$.

## 4 OFA Algorithms

In the following we give an algorithm for converting a DFA into an OFA that recognizes the same language as the DFA, while possibly processing fewer characters. Then we give an algorithm for matching using an OFA.

We can choose a different lookahead value for each state for which we build a trie. What value should we choose? Previously we showed that the lookahead value for state $q$ is $\leq finalDist(q)$ to prevent any final states from being skipped. Another important constraint is that the worst-case time and space to build a trie of a given lookahead is $O(|\Sigma|^{lookahead})$, which implies the time and space can quickly get unreasonable. So in practice we also set an upper limit based on available time and space. Also, assuming that we want to maximize the fraction of characters skipped, there is no reason to believe that a larger lookahead is always better, though there is some anecdotal evidence that it usually is.

Given these considerations, we give an algorithm which builds a trie and selects an appropriate lookahead. It works simply by growing a trie in steps from *lookahead*=1 to n by 1, stopping when we hit a resource constraint or when *lookahead* = *finalDist*(*q*). It is non-trivial to compute *finalDist*(*q*), and our algorithm relies on Theorem 14 to do so. It says that we have reached our max lookahead when the trie has a leaf whose state is a final state:

**Definition 13**: *reachedFinal*(*T1*) is true iff some leaf *T2* in *T1* has *state*(*T2*) $\in F$.

**Theorem 14:** *finalDist*(*q*) = min lookahead value for which *reachedFinal*(*T1*) is true when *isTrieForState*(*T1*, *q*, *lookahead*).

**Proof**: If *isTrieForState*(*T1*, *q*, *lookahead*), the set of states at leaf nodes of *T1* corresponds exactly to the set of states reachable in *lookahead* steps from *q*, which is also { $\delta(q, s)$ | $len(s) = lookahead$ }. If any of these is a final state, which is what *reachedFinal*(*T1*) computes, and if *lookahead* is the minimum such value $\geq 1$, then this is equivalent to *finalDist*(*q*) = *lookahead* by Definition 4. □

BuildTrie(*q*) returns a tuple <*T1*, *lookahead*> where *isTrieForState*(*T1*, *q*, *lookahead*), and $1 \leq$ *lookahead* $\leq$ *finalDist*(*q*) and *subRootCompressed*(*T1*).

```
function BuildTrie(q) returns <T1, lookahead>
  T1 = GrowTrie(trieLeaf(q))
  lookahead = 1
  while (not reachedFinal(T1)
         and BudgetAllowsBiggerTrie(q, T1, lookahead))
    T1 = GrowTrie(T1)
    lookahead = lookahead + 1
  endwhile
endfun
```
**Figure 8:** Algorithm for BuildTrie

To grow a trie, which means to increase the lookahead of a trie by 1, we rely on Theorem 17 that constructs a trie of lookahead $g+1$ from a trie of lookahead $g$.

**Definition 15**: *evolve*(*T1*, *c*) equals *T1* with each leaf *T2* updated via *state*(*T2*) = $\delta(state(T2), c)$.

**Lemma 16**: *selectState*(*evolve*(*T1*, *c*), *s*) = $\delta(selectState(T1, s), c)$.

**Proof**: *selectState*(*T1*) ignores the state value of leaves except when *T1* is a leaf. *T1* and *evolve*(*T1*, *c*) have the same structure, only the state value at leaves are different. If *selectState*(*T1*, *s*) returns the state value $q$ at a leaf, then $\delta(selectState(T1, s), c)$ is $\delta(q, c)$;

also, *selectState*(*evolve*(*T1*, *c*), *s*) will return the updated state value at the corresponding leaf, $\delta(q, c)$.  □

**Theorem 17**: If *isTrieForState*(*T2*, *q*, *lookahead*), and *T1* is a trie where *child*(*T1*, *c*) = *evolve*(*T2*, *c*) for all *c* in $\sum$, then *isTrieForState*(*T1*, *q*, *lookhead*+1)

**Proof**: For arbitrary *c*, and from Definition 1, Definition 2, Lemma 16, and the fact that *T1* is not a leaf, we have *selectState*(*T1*, *s*+*c*) = *selectState*(*child*(*T1*, *c*), *s*) = *selectState*(*evolve*(*T2*, *c*), *s*) = $\delta$(*selectState*(*T2*, *s*), *c*) = $\delta(\delta(q, s), c)$ = $\delta(q, s+c)$ which since *len*(*s*+*c*) = *lookahead*+1 means *isTrieForState*(*T1*, *q*, *lookahead*+1).  □

GrowTrie(*T1*) assumes *isTrieForState*(*T1*, *q*, *g*) for some *q* and lookahead *g*, and returns a *T2* satisfying *isTrieForState*(*T2*, *q*, *g*+1) and *subRootCompressed*(*T2*). *T2* has a lookahead 1 greater than *T1* and is compressed to the extent that no proper descendant node is compressible. Surprisingly, the function does not require the specific values of *q* and *g*.

```
function GrowTrie(T1) returns T2
  T2= trieNode()
  for c in ∑
    child(T2, c) = EvolveAndCompress(T1, c)
  endfor
endfun
```
**Figure 9**: Algorithm for GrowTrie

EvolveAndCompress(*T1*, *c*) returns a *T2* which is a copy of *T1* updated so that for all leaf *T3* in *T1*, with corresponding *T3'* in *T2*, *state*(*T3'*) = $\delta$(*state*(*T3*), *c*), and *compressed*(*T2*).

```
function EvolveAndCompress(T1, c) returns T2
  if (isLeaf(T1))
    T2= trieLeaf(δ(state(T1), c))
  else
    T2= trieNode()
    for (cx in ∑)
      child(T2, cx) = EvolveAndCompress(child(T1, cx), c)
    endfor
    if compressible(T2)
      T2= child(T2, c)
    endif
  endif
endfun
```
**Figure 10**: Algorithm for EvolveAndCompress

The budget function allows customizable control of the lookahead for each state. This function is used to limit the time and space of trie creation to fit the available budget. BudgetAllowsBiggerTrie(*q*, *T1*, *lookahead*) assumes *T1* satisfies *isTrieForState*(*T1*, *q*, *lookahead*) and *lookahead* < *finalDist*(*q*). It returns true iff we should grow *T1* to have *lookahead* = *lookahead* + 1. This version returns true if *lookeahead* is a reasonable value. Another reasonable implementation would return true when the worst-case memory size of the next bigger trie will be less than some maximum: *allowsBigger* = |$\sum$| * *memSize*(*T1*) < *MaxMemSizePerTrie*.

```
function BudgetAllowsBiggerTrie(q, T1, lookahead) returns allowsBigger
  allowsBigger = (lookahead < 12)
endfun
```
**Figure 11**: Algorithm for BudgetAllowsBiggerTrie

BuildOFA(DFA) builds an OFA M equivalent to a given DFA, following the process outlined in the previous section. It builds a trie for each state, and then assigns offsets to all transitions, and then links all tries into a single graph representing the OFA. BuildOFA assumes the following auxiliary data structures: *trieForState*[*q*] is a map mapping state *q* to its corresponding trie; *lookaheadForState*[*q*] is a map mapping state *q* to the lookahead of *trieForState*[*q*]. We use the notation $<c, d> =$ BuildTrie($q$) to denote the parallel assignment to variables $c$ and $d$ from the tuple $<T1, lookahead>$ returned by BuildTrie.

```
function BuildOFA(DFA) returns M
  for q in Q_DFA
    < trieForState[q], lookaheadForState[q] > = BuildTrie(q)
  endfor
  for q in Q_DFA
    AssignOffsets(trieForState[q])
  endfor
  for q in Q_DFA
    LinkTries(trieForState[q])
  endfor

  q0_M = trieForState[q0_DFA]
  F_M = { trieForState[q] | q in F_DFA }
  Q_M = {T1 | T1 is a trie node reachable from q0_M}
  for q in ( F_DFA + q0_DFA )
    Φ(trieForState[q]) = lookaheadForState[q] − 1
  endfor
  for qx in Q_M, c in ∑
    δ_M(qx, c) = child(qx, c)
    Θ(qx, c) = offset(qx, c)
  endfor
endfun
```
**Figure 12:** Algorithm for BuildOFA

Finally, we give an algorithm for processing an input string $s$ to locate all positions where the OFA M enters a final state.

```
function MatchUsingOFA(M, s) returns resultPositions
  resultPositions = if q0 ∈ F then {0} else { } fi
  curState = q0
  index = Φ(q0)
  while index < len(s)
    c = s[index]
    index = index + Θ(curState, c)
    curState = δ(curState, c)
    if (curState ∈ F)
      resultPositions = resultPositions ∪ (index – Φ(curState))
    endif
  endwhile
endfun
```
**Figure 13:** Algorithm for MatchUsingOFA

We point out that the main loop of MatchUsingOFA is nearly the same as the main loop of a routine for matching using a DFA: the DFA routine would have $\Theta(curState, c) = 1$ always, and $\Phi(curState) = 0$ always. This explains why we can do the complicated matching of an OFA nearly as fast as a DFA, even if no characters are skipped.

## 5 Complexity of BuildTrie

First we consider space complexity of BuildOFA(). The number of trie nodes and leaves in a trie of given lookahead is worst case $O(|\sum|^{lookahead})$ because it is a complete tree whose depth = *lookahead* with $|\sum|$ children per node and the worst case is that no node is compressible. The space of a single trie node and leaf is constant so $O(|\sum|^{lookahead})$ is also the space complexity of a single trie. We make a trie for every state: let $|Q|$ be the number of states in the DFA from which we construct our OFA. Every state has its own lookahead: let MaxLookahead be the largest lookahead for any state in Q. Then the best we can say is that the total space for tries is $O(|Q| |\sum|^{MaxLookahead})$. The space for the auxiliary data structures used in BuildOFA, the maps *trieForState*[] and *lookaheadForState*[], is linear in $|Q|$ and so their inclusion does not change the asymptotic complexity. The space to store an integer offset with each transition of the trie is proportional to the space to store the trie, since there is 1 transition to every node except for the root, and so its inclusion does not change the asymptotic complexity. The space to store the OFA is less than the space for the tries for each state, because we form the OFA by linking the tries for each state which removes the leaf nodes as a side effect. We conclude that the worst case space complexity of BuildOFA is $O(|Q| |\Sigma|^{MaxLookahead})$.

Next we consider the time complexity of BuildOFA(). We assume constant time insertion and lookup in our maps. To grow a trie from lookahead = $g-1$ to lookahead = $g$ requires constant time per trie node. Therefore a grow step to lookahead $g$ requires $O(|\Sigma|^g)$, and since we grow from lookahead 1..MaxLookahead in the worst case the total time to build a trie is given by the following sum:

$$O(|\Sigma|^{MaxLookahead}) + O(|\Sigma|^{MaxLookahead-1}) + O(|\Sigma|^{MaxLookahead-2}) + ... O(|\Sigma|^1)$$

which simplifies to $O(|\Sigma|^{MaxLookahead})$. Summing over all states the time to build all tries is worst-case $O(|Q| |\Sigma|^{MaxLookahead})$. AssignOffsets() and LinkTries() require a constant time per trie node or leaf, summed over all tries, so they do not change the asymptotic time after building the tries. Building the final OFA from the linked tries likewise requires constant time per trie node. We conclude the worst-case time complexity is $O(|Q| |\Sigma|^{MaxLookahead})$.

Because the complexity of BuildOFA(...) is exponential in $|\Sigma|$, we greatly benefit from techniques that minimize the alphabet size. Since most languages now support unicode strings which implies a very large alphabet, such techniques are in practice mandatory. The standard technique for minimizing the size of the alphabet is to split the alphabet into a set of equivalence classes, such that any two characters in the same equivalence class produce exactly the same transitions in the DFA or NFA. During compilation the primitive character classes used in the regular expression are translated into sets of equivalence classes. Then during matching each input character is mapped to the index of its equivalence class via an array lookup. See [8] and [9] for details.

Obviously, the worst-case matching time of MatchUsingOFA is linear in the size of the input, and requires space proportional to the number of ending match positions to store the results, which is worst case linear in the input size.

## 6 Test Results

We compare the space requirements, number of characters skipped, and elapsed time of the new algorithm MatchUsingOFA to previous algorithms. The previous algorithms include the "Forward-only DFA matcher", which is the classic algorithm for matching a regular expression, and the "LBwd-Pref backward and forward matcher", which is one of the linear time character skipping algorithms in [7]. All algorithms use the bit parallel implementation described in [7] to calculate states and transitions.

For English text searching, the "benglish-" patterns, we used [10], which is 405K and was lowercased and duplicated to be 10M.

For testing the DNA patterns, we used the *Haemophilus influenzae KW20 Rd* sequence from [11] which is 1.89MB long, modified to remove line breaks, and duplicated to be 10M.

Table 1 has the patterns we test against the English text and dna input, which are mostly the same patterns used in [7]. The *#EC* column abbreviates *# equivalence classes*; *Max LA* abbreviates *max lookahead*. The space column shows both the space used by the forward only algorithm and the additional space used by the OFA algorithm, in units of 32 bit words.

| Pattern Id | Pattern | #EC | Max LA | % Positions Matched | Space Forward-only + OFA |
| --- | --- | --- | --- | --- | --- |
| benglish1 | benjamin\|franklin | 12 | 8 | .01581% | 252 + 5K |
| benglish2 | benjamin\|franklin\|writing\|learning\|arithmetic | 17 | 7 | .03221% | 903 + 32K |
| benglish3 | [a-z][a-z0-9]*[a-z] | 3 | 2 | 58.766% | 25 + 29 |
| benglish3b | [a-z][a-z0-9]+[a-z] | 3 | 3 | 42.402% | 30 + 55 |
| benglish4 | benj.*min | 8 | 7 | .00522% | 140 + 38K |
| benglish5 | [a-z][a-z][a-z][a-z][a-z] | 2 | 5 | 20.067% | 28 + 59 |
| benglish6 | (benj.*min)\|(fra.*lin) | 12 | 6 | .01581% | 630 + 318K |
| benglish7 | ben(a\|(j\|a)*)min | 8 | 6 | .00522% | 100 + 3K |
| benglish8 | be.*ja.*in | 8 | 6 | .00968% | 160 + 84K |
| benglish9 | ben[jl]amin | 8 | 8 | .00522% | 100 + 1236 |
| benglish10 | (be\|fr)(nj\|an)(am\|kl)in | 8 | 8 | .01581% | 224 + 8K |
| benglish11 | 300 random words of length ≥ 4 drawn from the text. | 28 | 4 | .61743% | 284K + 4892K |
| dna1 | AC((A\|G)T)*A | 5 | 3 | 1.66422% | 49 + 127 |
| dna2 | AGT(TGACAG)*A | 5 | 4 | .31037% | 91 + 359 |
| dna3 | ((A\|CG)*\|(AC(T\|G))*)AG | 5 | 2 | 4.83337% | 63 + 111 |
| dna4 | AG(TC\|G)*TA | 5 | 4 | .41711% | 56 + 299 |
| dna5 | [ACG][ACG][ACG][ACG][ACG][ACG]T | 5 | 7 | 3.76455% | 45 + 259 |
| dna6 | TTTTTTTTTT[AG] | 3 | 11 | .00011% | 65 + 587 |
| dna7 | AGT[\u0000-\uffff]*AGT | 4 | 5 | 1.22507% | 54 + 2494 |

**Table 1**: Patterns tested

Note: The sample regular expressions use the subexpression ".*" which is shorthand for [^\n]*, namely 0 or more of any character excluding newline. In contrast, the subexpression "[\u0000-\uffff]*" indicates 0 or more of any unicode character with no exclusions.

Table 2 compares the percentage of characters processed, and percentage elapsed matching time, of the OFA character skipping algorithm and the LBwd-Pref character-skipping algorithm to the Forward-only algorithm. The percentages are calculated by dividing the number of characters processed, and the elapsed time, by the corresponding values for the Forward-only algorithm. Remember that the Forward-only algorithm examines each character of the input in order and is the best representative of the current state of the art.

To facilitate comparison we include in the rightmost colum the performance ratio, which is the inverse of the OFA percentage elapsed time. As an example, consider test "benglish1": the OFA algorithm processed only 18% of the characters, and required 30% of the time of the Forward-only algorithm; the LBwd-Pref processed 38% of the characters in 98% of the time of the Forward-only; and the OFA algorithm was 3.33 times faster than the Forward-only algorithm.

| Pattern Id | OFA % Chars Processed / Time | LBwd-Pref % Chars Processed / Time | Performance Ratio OFA vs. Forward-only |
| --- | --- | --- | --- |
| benglish1 | 18% / 30% | 38% / 98% | 3.33 |
| benglish2 | 23% / 40% | 47% / 118% | 2.50 |
| benglish3 | 97% / 99% | 100% / 126% | 1.01 |
| benglish3b | 91% / 86% | 100% / 124% | 1.16 |
| benglish4 | 56% / 40% | 98% / 93% | 2.50 |
| benglish5 | 72% / 77% | 130% / 154% | 1.30 |
| benglish6 | 66% / 53% | 193% / 188% | 1.88 |
| benglish7 | 22% / 35% | 45% / 113% | 2.86 |
| benglish8 | 84% / 63% | 193% / 176% | 1.59 |
| benglish9 | 17% / 27% | 35% / 83% | 3.70 |
| benglish10 | 18% / 30% | 38% / 95% | 3.33 |
| benglish11 | 89% / 140% | 149% / 258% | 0.71 |
| dna1 | 67% / 98% | 100% / 119% | 1.02 |
| dna2 | 58% / 78% | 131% / 271% | 1.28 |
| dna3 | 75% / 111% | 100% / 120% | 0.90 |
| dna4 | 63% / 68% | 137% / 244% | 1.47 |
| dna5 | 66% / 76% | 127% / 189% | 1.31 |
| dna6 | 21% / 30% | 45% / 100% | 3.33 |
| dna7 | 58% / 95% | 100% / 117% | 1.05 |

**Table 2:** Performance Results

Programs were written in Java and executed on a Java 7 runtime running on an Intel Core i7-3770T @ 2.50GHz, running a 64bit operating system (Windows 8). All times are in milliseconds, and measure the matching time only. Note that performing benchmarks on java code is very tricky, and even minor changes to the program can have dramatic and unexpected effects on runtime performance. See [12] and [13] for a good overview.

OFA outperformed LBwd-Pref in every case in both metrics, characters processed and elapsed time. OFA was faster than Forward-only in 17/19 cases, and was at least 2.5 times faster in 7/19 cases.

We found the times for OFA and its main competitor, Forward-only, could best be explained with the following regression equation:

*time* * 10^6 = 4.88852 * *NumMatches*
    + 4.16018 * *ForwardOnlyCharsRead* + 4.63247 * *OFACharsRead*
    + 3.39653 * *PatternSize*

This regression has an $R^2$ of .98. Note that for the OFA algorithm, *ForwardOnlyCharsRead* will always be 0, and for the Forward-only algorithm the *OFACharsRead* will always be 0. The equation is built on the observation that the matching algorithm's time complexity is a linear function of the number of characters processed plus the number of matches. The equation has a zero constant value since processing zero characters takes essentially no time. Since both algorithms have exactly the same code for processing matches, and since even skipping algorithms have to process every match, the factor based on the number of matches is the same for both.

The factor *PatternSize* was a surprise, though its magnitude is only suggestive since only benglish11 and benglish6 have significant data sizes and these two cases are insufficient to provide accurate statistics. In testing benglish11 we were surprised to find that the OFA algorithm took 40% more time than Forward-only, despite processing 10% fewer characters. We found that this was explained by the much larger data size of OFA compared with Forward-only, which we confirmed by rewriting Foward-only data so that its data was spread out over a comparable data size and noting a significant slowdown. We hypothesize that this behavior is due to caching behavior of the computer system. As a result, it is quite unlikely that the factor is linear as suggested by the equation; instead, the slowdown is most likely a complex function of the size of the data, the system cache size, and the locality of the matching among the DFA states.

A few of the tests, such as benglish6, showed an elapsed time percentage that was smaller than the characters processed percentage. This is surprising but the fact that a similar effect was recorded for both OFA and LBwd-Pref suggests that it was a slowdown in the denominator of the percentage, i.e. an increase in the elapsed time for the forward-only algorithm that is responsible, perhaps due to a java benchmarking quirk or caching issue.

The OFA algorithm is slightly more complicated than the Forward-only algorithm. The regression quantifies this extra complexity as costing about 12% per character processed. As a result, we would expect that OFA would outperform the forward-only algorithm if it skipped 15% or more of the input.

## 7 Conclusion

We presented an algorithm for converting a DFA into an equivalent Offsetting Finite Automaton (OFA), and we gave a matching algorithm using an OFA. The algorithm never processes a character of the input more than once, and may skip a substantial amount of the input. We showed that the algorithm can provide substantial speedups in matching sample patterns. The main drawback of the algorithm is increased preprocessing time before matching begins. We showed how to include a "budget" function to limit preprocessing time and/or space to the available budget. The main limitation of the algorithm is that the maximum number of characters we can avoid processing from any state is one less than the length of the shortest string matched by the regular expression.

### 7.1 Future Extensions

If the performance metric is maximizing the percentage of skipped characters, it is not clear what lookahead value is optimal. Anecdotally it appears that larger lookahead values are generally better, but in general this outcome is not guaranteed. Every time we grow a trie's lookahead by 1, we increase the path from the root to every leaf by 1, while increasing the lookahead amount by 1, which in the absence of any compression would decrease the percentage of skipped characters. We hope that enough nodes of the trie then get compressed to more than compensate for this increase. For example, a trie of lookahead 6 might compute the state 6 characters ahead while skipping 2 of the characters on average, or 33%. A trie for the same state of lookahead 7 computes the state 7 characters ahead but without additional compression will still skip 2 of the characters on average, so the percentage of characters skipped would decrease to $2/7 = 28\%$. Better criteria for selecting a good lookahead would be valuable.

Since matching sets of strings is a subset of the regular expression matching problem, the OFA algorithm can theoretically replace all string matching algorithms that match sets of strings. Of course, existing algorithms are significantly faster in preparing the data structures for matching and thus are preferred for many situations. It would be helpful to clarify the situations in which the algorithm reported here can replace or augment the myriad sublinear algorithms for matching sets of strings.

Although our trie are tree structures, there is no difficulty in storing them as directed acyclic graphs, so that common subtrees are shared among trie. This may save substantial time and space while building tries, though it does not effect matching time, since the savings are exponential as lookahead increases. Therefore it would be interesting to study the frequency of common subtrees among trie.

Lazy DFA computation is extremely useful for getting the best matching performance and the smallest precompile time. Finding techniques to build an OFA "lazily" would also be very helpful.